\magnification=1200
\hsize =36true pc\vsize = 48true pc
\hoffset=.375 true in
\voffset=.5true in

\font\mysmall=cmr8 at 8 pt
\font\eightit=cmti8

\def\g{\bf G}

\def\q{\bf Q}

\def\m{\bf M}
\def\p{\bf P}
\def\r{\bf R}
\def\v{\bf V}

\def\hh{\cal H}
\def\kk{\cal K}

\def\uu{\cal U}

\def\uu{\cal U}

\def\picture #1 by #2 (#3){
		\vbox to #2{
		\hrule width #1 height 0pt depth 0pt
		\vfill
		\special {picture #3}}}
\font\teneufm eufm10
\font\seveneufm eufm7
\font\fiveeufm eufm5
\newfam\eufm
\textfont\eufm\teneufm
\scriptfont\eufm\seveneufm
\scriptscriptfont\eufm\fiveeufm
\def\frak#1{{\fam\eufm\teneufm#1}}
 
\def\picture #1 by #2 (#3){
		\vbox to #2{
		\hrule width #1 height 0pt depth 0pt
		\vfill
		\special {picture #3}}}
\centerline {\bf Structure, classification, and conformal symmetry,}
\bigskip
\centerline {\bf of elementary particles over non-archimedean space-time}
\vskip0.5 true in\noindent
\centerline {\bf V. S. Varadarajan and Jukka Virtanen}
\vskip 1 true in\noindent
{\mysmall
\centerline {ABSTRACT}
\vskip 0.5 true in\noindent
It is known that no length or time measurements are possible in sub-Planckian regions of spacetime. The Volovich hypothesis postulates that the micro-geometry of spacetime may therefore be assumed to be non-archimedean. In this letter, the consequences of this hypothesis for the structure, classification, and conformal symmetry of elementary particles, when spacetime is a flat space over a non-archimedean field such as the $p$-adic numbers, is explored. Both the Poincar\'e and Galilean groups are treated. The results are based on a new variant of the Mackey machine for projective unitary representations of semidirect product groups which are locally compact and second countable. Conformal spacetime is constructed over $p$-adic fields and the impossibility of conformal symmetry of massive and eventually massive particles is proved.
\vskip 0.3 true in\noindent
\vskip 0.3 true in\noindent
}
\vskip 0.3 true in\noindent
{\bf 1. Introduction.\/} In the 1970's many physicists, concerned about the divergences in quantum field theories, started exploring the micro-structure of space-time itself as a possible source of these problems. In particular, Beltrametti and his collaborators proposed the idea in [3] [4] [5] that the geometry of space-time might be based on a non-archimedean, or even a finite, field and examined some of the consequences of this hypothesis. But the idea did not really take off until Volovich proposed in 1987 [20] that world geometry at sub-Planckian regimes might be non-archimedean. The reasoning behind this hypothesis is that no measurements  are possible at such ultra-small distances and time scales, due to the interplay between general relativity and quantum theory. Indeed, the Planck scale emerges naturally when one identifies the Schwarzchild radius and the Compton wave length. Since impossibility of measurements automatically forbids also comparisons between different lengths and also different times, the Volovich hypothesis is very natural. Since no single prime can be given distinguished status, it is even more natural to see if one could really work with an {\it adelic\/} geometry as the basis for space-time. Such an idea was first proposed by Manin [13]. A huge number of articles have appeared since then, exploring these and related themes. For  a definitive survey and a very inclusive set of references see the very recent article by Dragovich et al [7]. In this letter we describe some results that have come out of our examination of the consequences of the non-archimedean hypothesis for the structure and classification of elementary particles. We consider both the Poincar\'e and the Galilean groups. Our methods apply to both the local and adelic geometries but in this note the main emphasis is on local non-archimedean geometry. Details will appear in a later publication.
\medskip
One knows that (see [18]) that the symmetry of a quantum system with respect to a group $G$, locally compact and second countable (lcsc), may be expressed by a projective unitary representation (PUR), either of $G$ or of a subgroup of index $2$ in $G$, in the Hilbert space of quantum states; this PUR may be lifted to an ordinary unitary representation (UR) of a suitable topological central extension (TCE) of the group by the circle group $T$. The  PUIRs (=irreducible PURs) of $G$ then classify the elementary particles with $G$-symmetry, with or without selection rules or sectors (real mass, positive energy, etc). In the supersymmetric world, when $G$ is a {\it real\/} super Lie group and we consider only {\it ordinary unitary representations\/}, the classification of superparticles, long understood by the physicists heuristically, was carried out in [6] (see also [17]). The extension of supersymmetry to non-archimedean or adelic world geometry is an open problem.
\medskip
Going beyond particle classification is the construction of quantum fields over non-archimedean spacetime. The most penetrating work on these issues so far is [9].
\medskip
Returning to particle classification, Wigner [21], proved that all PURs of the connected {\it real\/} Poincar\'e group $P$ lift to URs of the simply connected covering group $P^\ast={\r}^{1,3}\times '{\rm Spin}({\r}^{1,3})$ of $P$ where $\times '$ denotes semidirect product. In other words, $P^\ast$ is already the {\it universal\/} TCE of the Poincar\'e group. Thus particles with $P$-symmetry are classified by UIRs of $P^\ast$. Now for any semidirect product the Mackey machine is applicable; and for $P^\ast$ it just gives the Wigner theory.\medskip
The situation over a disconnected field is more complicated. To explain this we need a little terminology. Let $k$ be a field of arbitrary characteristic. If ${\m}$ is a linear algebraic group defined over $k$ and $r$ is an extension field of $k$, we write $M(r)$ for the group of $r$-points of ${\m}$. If $k$ is a locally compact field then $M(k)$ is a lcsc group and one can ask whether it has a universal TCE so that PURs of $M(k)$ can be treated as URs of this universal extension. However not all lcsc groups have universal TCEs; it is necessary for example that their  commutator subgroups should be dense in them. Over a non-archimedean local field, the commutator subgroups of $M(k)$ are often open and closed subgroups of $M(k)$, and it is generally a very delicate procedure to verify whether they are equal to $M(k)$. So it is preferable to work with the PURs of $M(k)$ itself, rather than look for TCEs of $M(k)$. We note that for $k$ non-archimedean local, the groups $M(k)$ are totally disconnected.
\medskip
Number theorists have long been interested in URs of groups $M(k)$ for {\it simple\/} groups ${\m}$. In physics groups with radical appear to be important and so it is worthwhile to study URs of these groups as well.
\medskip
Fix a non-archimedean local field $k$ of characteristic $\not=2$. Let ${\v}$ be an isotropic  quadratic vector space over $k$; this means that ${\v}$ has a non-degenerate quadratic form defined over $k$ which has null vectors over $k$. Then  we have the algebraic groups ${\g}={\rm SO}({\v})$ and its two-fold cover ${\g}_{spin}={\rm Spin} ({\v})$. We thus have correspondingly the Poincar\'e groups ${\p}={\v}\times ^\prime {\g}$ and ${\p}_{spin}={\v}\times ^\prime {\g}_{spin}$. Write $V, G$ and $G_{spin}$ for the groups of $k$-points of ${\v}, {\g}$,  and ${\g}_{spin}$ respectively, and $P, P_{spin}$ for the respective groups of $k$-points of ${\p}, {\p}_{spin}$. Now $G_{spin}$ and $P_{spin}$ {\it do have\/} TCE's; for the spin groups this is a consequence of the work of Moore [14] and Prasad and Raghunathan [16] and for the corresponding Poincar\'e groups, of the work of Varadarajan [19]. Moreover, if $G_{spin}^\ast$ is the universal TCE of $G_{spin}$, it is shown in [19] that the universal TCE $P_{spin}^\ast$ of $P_{spin}$ is given by $P_{spin}^\ast=V\times ^\prime G_{spin}^\ast$. So all PURs of $P_{spin}$ lift to URs of $P_{spin}^\ast$, and since $P^\ast$ is a semidirect product, the Mackey-Wigner theory is applicable. We are thus in the same situation as in the real case and there are no fundamental obstacles to the classification of the particles (=irreducible PURs) with $P_{spin}$-symmetry.
\medskip
However the natural maps  $G_{spin}\longrightarrow G$ and $P_{spin}\longrightarrow P$ are {\it not} surjective (even though they are surjective over the algebraic closure of $k$), and so replacing the orthogonal group $G$ by the spin group $G_{spin}$ leads to a loss of information. So we work with the orthogonal group rather than the spin group. To illustrate this point, let $G={\rm SL}(2, {\q}_p)$. The adjoint representation exhibits $G$ as the spin group corresponding to the quadratic vector space $\frak g$ which is the Lie algebra of $G$ equipped with the Killing form. The adjoint map $G\longrightarrow G_1={\rm SO}(\frak g)$ is the spin covering for ${\rm SO}(\frak g)$ but this is {\it not surjective\/}; in the standard basis
$$
X=\pmatrix {0&1\cr 0&0\cr}, \quad H=\pmatrix {1&0\cr 0&-1\cr}, \quad Y=\pmatrix {0&0\cr 1&0\cr}
$$
the spin covering map is
$$
\pmatrix{a&b\cr c&d\cr}\longmapsto \pmatrix {a^2&-2ab&-b^2\cr -ac&2cd&bd\cr -c^2&-(ac+bd)&d^2\cr}.
$$
The matrix
$$
\pmatrix {\alpha&0&0\cr 0&1&0\cr 0&0&\alpha^{-1}\cr}
$$
is in ${\rm SO}(\frak g)$; if it is the image of $\pmatrix {a&b\cr c&d\cr}$, then $b=c=0$ and $\alpha=a^2$, so that unless $\alpha \in {{\q}_p^\times}^2$, this will not happen.
\medskip
The group $P$ is still a semidirect product but we are now required to determine its irreducible PURs.  This means that we must determine its multipliers and then, to each multiplier $m$, find the irreducible $m$-representations. It turns out that there is a very nice variant of the Mackey machine for $m$-representations of a seimidirect product that allows us to do this. In this letter we describe this variant, which appears not to have been noticed in the literature, and then apply it to the Poincar\'e and Galilean groups over a non-archimedean local field. The variant is formulated in the framework of locally compact groups and so is applicable to adelic geometries as well, but here we restrict ourselves to the local case. We assume that the reader is familiar with the basic ideas of PURs, multipliers, and so on; see [18] [12]. For any locally compact second countable (lcsc) group $G$ we write $Z^2(G)$ for the group of its multipliers and $H^2(G)$ the quotient of $Z^2(G)$ by the subgroup of trivial multipliers. If $G$ is totally disconnected, every multiplier is equivalent o a continuous one, and in fact, the Borel cohomology group is canonically isomorphic to the continuous cohomology group [19].
\bigskip\noindent
{\bf 2. Multipliers and PURs for semidirect products.\/} Let $H = A \times^\prime G$ where $A$ and $G$ are lcsc groups and $A$ is abelian. Let $A^*$ be the
character group of $A$. We define a $1$-cocycle for $G$ with coefficients in $A^*$ as a Borel map $f(G \longrightarrow A^*)$ such that $$f(gg') = f(g) + g [ f(g') ]
\;\; (g,g' \in G)$$ or equivalently that $g \mapsto
(f(g),g)$ is a Borel homomorphism of $G$ into the semidirect product
$A^* \times ^\prime G$, so that all 1-cocycles are continuous. The
abelian group of continuous $1$-cocycles is $Z^1(G,A^*)$ and the
coboundaries are cocycles of the form $g \mapsto g [\chi] - \chi$
for some $\chi \in A^*$. These form a subgroup $B^1(G,A^*)$
of $Z^1(G,A^*)$ to give the cohomology group $H^1(G,A^*) =
Z^1(G,A^*) \slash B^1(G,A^*)$.
\medskip
Let $M_A(H)$ the group of multipliers on $H$ that are trivial
when restricted to $A \times A$.  Let $H^2_A(H)$ denote its image in
$H^2(H)$.  Let $M'_A(H)$ be the group of multipliers $m$ for $H$
with $m\vert_{A \times A} = m\vert_{A \times G} = 1.$
\medskip
From  [19] [11] we find that any element in $M_A(H)$ is
equivalent to one in $M'_A(H)$. If $m \in M'_A(H), m_G =
m\vert_{G \times G}$, and $\theta_m(g^{-1})(a') = m(g,a')$, then $\theta_m\in Z^1(G, A^\ast)$, and $m
\mapsto(m_G, \theta_m)$ is an isomorphism $M'_A(H) \simeq Z^2(G)
\times Z^1(G,A^*)$ which is well defined in cohomology and gives the
isomorphisms $H^2_A(H) \simeq H^2(G) \times H^1(G,A^*)$. It follows from this that if $n \in Z^2(G)$ and $\theta\in Z^1(G,A^*)$ are given then one may define
$m \in M_A'(H)$ by $m(ag,a'g') =n(g,g')\theta(g^{-1})(a')$. If $n=1$ then $m(ag,a'g') =
\theta(g^{-1})(a')$. If $H^1(G,A^*)=0$ every multiplier of
$H$ is equivalent to the lift to $H$ of a multiplier for $G$.

\medskip
Let $G$ be a lcsc group. Let $X$ be a $G$-space that is also a standard Borel space. Let ${\hh}$ be a separable Hilbert space and ${\uu}$ the
unitary group of ${\hh}$. An $m$-representation of $G$ is a Borel map $U$ of $G$ into the unitary group of a separable Hilbert space such that $U(e)=1$ and $U(g)U(g') = m(g,g') U(gg')$. An {\it $m$-system of imprimitivity} based on $X$ is a pair $(U,P)$, where $P(E
\rightarrow P_E)$ is a projection valued measure on the class of
Borel subsets of $X$, the projections being defined in
${\hh}$, and $U$ is an $m$-representation of $G$ in
${\hh}$ such that
$$
U(g)P(E)U(g)^{-1} = P(g[E])
$$
for all $g\in G$ and Borel $E\subset X$. Let $X$ be a transitive $G$-space. We fix some $x_0\in X$ and let $G_0$ be the stabilizer of $x_0$ in $G$,
so that $X \simeq G \slash G_0$. We will also fix a multiplier $m$
for $G$ and let $m_{G_0} = m\vert_{G_0 \times G_0}$. Then Mackey's technique of unitarizing projective representations by going to a suitable TCE leads to a natural one to one correspondence between
the $m_{G_0}$-representations $\mu$ of $G_0$ and $m$-systems of
imprimitivity $(U,P)$ of $G$  based on $X$. Under this
correspondence we have a ring isomorphism of the commuting ring of
$\mu$ with that of $(U,P)$, so that irreducible $\mu$ correspond to irreducible $(U, P)$.
\medskip
In order to use this point of view in determining PURs of semidirect products we shall now introduce certain new actions of $G$ on $A^\ast$ defined by cocycles in $Z(G, A^\ast)$. If $\theta:G \rightarrow A^*$ is a continuous map with $\theta (1)=0$, then, defining $g\{\chi \} = g[\chi] + \theta(g)$, for $g \in G, \chi \in
A^*$, it is easy to see that $g:\chi \mapsto g\{\chi\}$ defines an action of $G$ on $A^*$ if and only if $\theta \in Z^1(G,A^*)$. This action depends on the choice of the cocycle $\theta \in Z^1(G,A^*)$, so we write it as $g_\theta\{\chi\}$. The actions defined by $\theta$ and $\theta'$ are equivalent in the following  sense: if $\theta'(g)=\theta(g)+g[\xi]-\xi$ where $\xi\in A^\ast$, then $g_{\theta'}=\tau ^{-1}\circ g_{\theta}\circ \tau$ where $\tau$ is the translation by $\xi$ in $A^\ast$.  The action $g_\theta:\chi \mapsto g_\theta\{\chi\}$ is called {\it the affine action\/} of $G$ on $A^*$ determined by $\theta$.
\medskip
The following theorem now shows how the $m$-representations of $H$
correspond to $m_G$-systems of imprimitivity on $A^*$ where the
action of $G$ on $A^*$ is given by the affine action.
\bigskip\noindent
{\bf Theorem 1.\/} {\it Fix $\theta \in Z^1(G,A^*)$ and $m \in M_A'(H), \; m \simeq (m_G,\theta)$. Then there is a natural bijection between $m$-representations $V$ of $H=A
\times^\prime G$ and $m_G$-systems of imprimitivity $(U,P)$ on $A^*$ for
the affine action $g_\theta:\chi \mapsto g\{\chi\} =
g_\theta\{\chi\}$ defined by $\theta$. The bijection is given by

$$V(ag) = U(a)U(g), \;\;\; U(a) = \int_{A^*}<a,\chi> dP(\chi).$$
}
\medskip
We now obtain the basic theorem of irreducible $m$-representations
of $H$.
\bigskip\noindent
{\bf Theorem 2.\/} {\it Fix $\chi \in A^*$, $m \simeq(m_G, \theta)$.  Then there is a natural
bijection between irreducible $m$-representations $V$ of $H=A
\times^\prime G$ with $Spec(V) \subset G\{\chi \}$ $($the orbit of $\chi$
under the affine action$)$ and irreducible $m_G$-representations of
$G_\chi$, the stabilizer of $\chi$ in $G$ for the affine action.  If
the affine action is regular, every irreducible $m$-representation
of $H$, up to unitary equivalence, is obtained by this procedure.\/}
\bigskip\noindent
{\bf Corollary.\/} {\it Suppose $H^1(G,A^*)=0$. Then we can take
$\theta(g)=1$ and $m(ag,a'g')= m_G(g,g')$.  In this case, the
affine action reduces to the ordinary action.\/}
\bigskip\noindent
{\bf Remark.\/} It follows easily from the relationship between the affine actions defined by two cocycles $\theta, \theta'$ described earlier that the classes of PURs defined by $\theta$ and $\theta'$ are equivalent.
\bigskip\noindent
{\bf 3. The Poincar\'e group over an arbitrary field and particle structure and classification over the $p$-adic numbers.\/} Let $V$ be  finite dimensional, isotropic quadratic vector space over a field $k$ of ch $\neq 2$. Let $G=SO(V)$ be the group of $k$-points of the corresponding orthogonal group preserving the quadratic form. By the $k$-Poincar\'{e} group we shall
mean the group
$$
P_k = V \times^\prime G.
$$
It is the group of $k$-points of the corresponding algebraic group which is defined over $k$.
\medskip
From now on we assume that $k$ is a non-archimedean local field. The requirement of Minkowki signature does not make sense over $k$. Instead we fix the {\it Witt class\/} (see [10]). The results below do not depend on the Witt class. We write $V^\prime$ for the algebraic dual of $V$. With its $k$-topology it becomes isomorphic with $V^\ast$ as a $G$-module.
\medskip
Since $V$ is isotropic, the cohomology $H^1(G, V^\ast)=0$ [19]. Hence the PUIRs of $P_k$ can be obtained by the theorems of \S 2. They are classified by the orbits of $G$ in $V^\prime$. The orbits are: the level sets of the quadratic form when the value of the quadratic form (called the {\it mass\/}) is non-zero; the level set of zero with the origin deleted; and the singleton consisting of $0$. These are referred to as {\it massive}, {\it massless\/}, and {\it trivial massless\/} respectively. The orbit action is regular by a theorem of Effros [8] since all orbits are either closed or open in their closure.
\bigskip\noindent
{\bf Theorem 1.\/} {\it Let $P_k=V \times^\prime G$ be the $k$-Poincar\'{e} group.  Fix
$p \in V^\prime$, $m_0$ be a multiplier of $G$, and let $m$ its
lift to $P_k$. Then there is a natural bijection between irreducible
$m$-representations of $P_k=V \times^\prime G$ with $Spec(V) \subset
G[p]$, the orbit of $p$ under the natural action of
$G$, and irreducible $m_{0p}$-representations of $G_p$, the
stabilizer of $p$ in $G$, $m_{0p}$ being the restriction of $m_0$ to $G_p$.  Every PUIR of $P_k$, up to unitary
equivalence, is obtained by this procedure. \/}
\bigskip\noindent
{\bf Remark 1.\/} Let $X =G[p]$, and let
$\lambda$ be a $\sigma$-finite quasi-invariant measure on $X$ for the action of
$G$. Then, for any irreducible $m_{0p}$-representation $\mu$ of
$G_p$ in the Hilbert space ${\kk}$, the corresponding
$m$-representation $U$ acts on $L^2(X,{\kk}, \lambda)$ and has the
following form:
$$
(U(ag)f)(q) = \psi(<a,q>) \rho_g (g^{-1}[q])^{1/2})
\delta(g,g^{-1}[q])f(g^{-1}[q])
$$
where $\delta$ is
any strict $m_{0p}$-cocyle for $(G, X)$ with values in
${\uu}$, the unitary group of ${\kk}$, such that
$\delta(g, q) = \mu(g), \; g\in G_p$.
\bigskip\noindent
{\bf Remark 2.\/} This theorem shows that the elementary particles over $k$
have a richer structure than in the real case.  The PUIRs are still classified by mass,
but for a given mass, by the PUIRs of the stabilizer in $G$ of a
point in that mass orbit. Unlike the real case we cannot replace the PUIRs of the little groups by URs of a single TCE of these groups. The determination of all the multipliers of the little groups is not treated here.
\bigskip\noindent
{\bf 4. Galilean group and Galilean particles.\/} Here spacetime $V=k^{r+1}$ has the decomposition into space and time: $V=V_0\oplus V_1$ where $V_0=k^r, V_1=k$. The Galilean group is the semi direct product $G=V\times 'R$ where $R$ itself is the semi direct product of rotations and boosts. Thus $V_0$ is a quadratic vector space. We set $R_0={\rm SO}(V_0)$ at first and set $R=V_0\times ' R_0$. The action of $G$ is defined by
$$
r=((u, \eta), (v, W)) : (x, t)\longmapsto (Wx+tv+u, t+\eta).
$$
We write $( {\cdot} , {\cdot})$ for the bilinear form on $V_0$. The dual $V'$ consists of pairs $(\xi, t)$ with duality $\langle (\xi, t), (u, \eta)\rangle=(\xi, u)+t\eta$. The actions of the group $R_0$ on $V$ and $V'$ are given by
$$
(v, W) : (u, \eta)\longmapsto (Wu+\eta v,\eta),\qquad
(v, W) : (\xi, t)\longmapsto (W\xi, t-(W\xi, v)).
$$
\medskip
Let
$$
\theta_\tau (v, W)=(-2\tau v, -\tau (v, v))\qquad (\tau\in k, (v, W)\in R).
$$
The $\theta_\tau$ are in $Z^1(R, V^\prime)$ and $\tau\longmapsto [\theta_\tau]$ is an isomorphism of $k$ with $H^1(R, V^\prime)$. Let $n_0$ be a multiplier for $R_0$ and let $n$ be the lift to $G$ of $n_0$ via the composition of the maps $G\longrightarrow R$ and $R\longrightarrow R_0$. Define $m_{n_0, \tau}$ by
$$
m_{n_0, \tau}(r, r')=n_0((v, W), (v', W'))\psi (-2\tau (v, Wu')-\eta'(v,v))
$$
for $r=((u, \eta), (v, W)), r'=((u', \eta'), (v', W'))\in G$. Then it follows from [19] that
$$
(n_0, \tau)\longmapsto nm_{n_0, \tau}
$$
gives an isomorphism of $H^2(R_0)\times k$ with $H^2(G)$.
\medskip
From this we can determine the Galilean particles. The analysis is somewhat involved and we just give the highlights. First of all the representations corresponding to $\tau=0$ are ordinary UIRs and are rejected as in [18]. Fix now $\tau\not=0$. The affine action corresponding to the cocycle $\theta_\tau$ is given by
$$
(v, W) : (\xi, t)\longmapsto (W\xi+2\tau v, t-(W\xi, v)-\tau (v,v)).
$$
It is an easy calculation that the function
$$
M : (\xi, t)\longmapsto (\xi, \xi)+4\tau t
$$
is invariant and maps onto $k$, since $M((0, a/4\tau))=a$. If $M((\xi, t))=a$ the element $(\xi/2\tau, I)$ of $R$ sends $(0, a/4\tau)$ to $(\xi, t)$. Hence the sets $M_a$ where $M$ takes the value $a$ are orbits for the affine action. The stabilizer in $R$ of $(0, a/4\tau)$ is just $R_0$. Hence for a given $a$ the $m_{n_0, \tau}$-representations are parametrized by the $n_0$-representations of $R_0$.
\medskip
However a little more analysis as in [18] reveals that {\it for different $a$ all these representations are projectively the same.\/} The map
$$
\xi \longmapsto \bigg (\xi, (1/4\tau)(a-(\xi, \xi)\bigg)
$$
is a bijection of $V_0^\prime$ with the orbit $M_a$. The action of $R$ on $M_a$ becomes the action
$$
\xi\longmapsto W\xi+2\tau v
$$
under this bijection and so Lebesgue measure is invariant. {\it The parameter $a$ has disappeared in the action.\/} In the Hilbert space of the corresponding representation, the spacetime translation $(u, \eta)$ acts as multiplication by
$$
\psi \bigg ((u, \xi)+{\eta (a-(\xi, \xi)\over 4\tau}\bigg ).
$$
The factor
$$
\psi \bigg ({\eta a\over 4\tau }\bigg)
$$
pulls out and is independent of the variable $\xi$. Hence it is a phase factor and can be omitted. {\it The resulting projective representation is thus independent of $a$.\/} Hence all these representations represent a single particle. The true parameters are $\tau (\not=0)$ and the projective representations $\mu$ of $R_0$. We interpret $\tau$ as the {\it Schr\"odinger mass\/}, and $\mu$ as the {\it spin\/}.
\bigskip\noindent
{\bf 5. Conformal compactification of $p$-adic spacetime and conformal symmetry of $p$-adic Poincar\'e particles.\/} Over the reals the Poincar\'e group of the Minkowski space ${\r}^{1, n}$ can be imbedded in the {\it conformal group\/} ${\rm SO}(2, n+1)$ in such a way that the space-time is dense and open in a compact homogeneous space for the conformal group. This can be done over any field $k$ of characteristic $\not=2$.
\medskip
Let $k$ be a field of characteristic $\not=2$ with algebraic closure $\bar k$, $V$ a quadratic vector space over $k$, $\bar V=\bar k\otimes V$, and $P_k$ (resp. $P_{\bar k}$) the $k$-Poincar\' group of $V$ (resp. the $\bar k$-Poincar\'e group of $\bar V$). Let $V_0=V\oplus U$ where $U$ is a quadratic vector space with a basis $p, q$ such that $(p,p)=(q,q)=0, (p,q)=1$. We define $\bar V_0=\bar k\otimes V_0, \bar U=\bar k\otimes U$. Then $H={\rm SO}(\bar V_0)$ is an algebraic group defined over $k$. We write $H(k)$ for its group of $k$-points.
\bigskip\noindent
{\bf Theorem 1.\/} {\it The group $P_{\bar k}$ is isomorphic, as an algebraic group over $k$ to the stabilizer $H_p$ of $p$ in $H$. The isomorphism is defined over $k$ and gives an isomorphism of $P_k$ with $H_p(k)$, the stabilizer of $p$ in $H(k)$. If $\dim (V)\ge 5$, then all $k$-imbeddings of $P_{\bar k}$ in $H$ are conjugate over $H(k)$.\/}
\bigskip\noindent
{\bf Remark 1.\/} Writing $V_0$ as $kp\oplus kq\oplus V$ the imbedding is given (in block matrix form) by
$$
(t,R)\longmapsto \pmatrix {1&-{(t,t)\over 2}&e(t, R)\cr 0&1&0\cr 0&t&R\cr}
$$
where $e(t, R)\in {\rm Hom} (V, kp)$ is the map $v\longmapsto (t, Rv)p$ and $R\in {\rm SO}(V)$.
\bigskip\noindent
{\bf Remark 2.\/} The conjugacy of the imbeddings can be proved using the theory of parabolic subgroups of $H$. But a direct proof using only the basics of the theory of linear algebraic groups is possible.
\medskip
Let $\Omega$ be the cone of null vectors in $V_0$ and $[\Omega]$ its image in projective space. Let $A_p = \{a \in \Omega \vert (p,a) \neq 0 \}$. Then $a = \alpha p + \beta q + w$, where $w\in V$, and $\beta \neq 0$. Taking $\beta=1$ does not change the image $[a]$ of $a$ in projective space, and then $\alpha = -(w,w)/2$ so that $[a]$ is given by $[-(w,w)/2 : 1 : w]$. Thus $[a]$ is entirely determined by $w$. Thus $J: w \mapsto [-(w,w)/2 : 1 : w]$ is a bijection of $V$ with the image $[A_p]$ of $A_p$ in projective space.  Then we have the following theorem.
\bigskip\noindent
{\bf Theorem2.\/} {\it There is a natural conformal structure on $[\Omega]$, and the group $H(k)$ acts transitively on $[\Omega]$. Moreover $[A_p]$ is a Zariski open dense subset of $[\Omega]$ stable under $H_p$, and the imbedding $J$ intertwines the action of the Poincar\'e group $P_k$ with that of $H_p$ on $[\Omega]$.\/}
\bigskip
When $k$ is a local field, $[\Omega]$ is compact and so we have a compactification of space-time into $[\Omega]$. For this reason it is natural to call $[\Omega]$ the {\it conformal space-time\/} over $k$.
\bigskip\noindent
{\bf Partial conformal group.\/} The {\it partial conformal group\/} is defined as the stabilizer of $A_{[p]}$ in the conformal group. We denote it by
$\widetilde P(W, k)$. We have
$$
P(W, k)\simeq H_p\subset \widetilde P(W, k).
$$
It can be shown that $\widetilde P(W, k)$ is the stabilizer of the line $kp$ in the conformal group. It is isomorphic to the subgroup of ${\rm SO}(V, k)$ of matrices of the form
$$
\pmatrix {c&{-c(t, t)\over 2}&ce(t, R)\cr 0&c^{-1}&0\cr 0&t&R\cr}
\quad (c\in k^\times, t\in W, R\in {\rm SO}(W, k)).
$$
In particular
$$
\widetilde P(W, k)\simeq P(W, k)\times ^\prime k^\times
$$
where $c\in k^\times$ commutes with ${\rm SO}(W, k)$ and acts as a dilatation, namely, multiplication by $c$ on $W$.
\medskip
The conformal group in general will move points of spacetime into the infinite part $[\Omega]\setminus A_{[p]}$. It is only the Poincar\'e group extended by the dilatations that will leave spacetime invariant.
\bigskip\noindent
{\bf Partial and full conformal symmetry.\/} An elementary particle or the corresponding UIR {\it has partial conformal symmetry if it extends to a UR of $\widetilde P(W, k)$. An elementary particle or the corresponding UIR {\it has full conformal symmetry if it extends to a UR of ${\rm SO}(V, k)$.\/} It is natural to ask which particles, if any, have partial or full conformal symmetry.\/}
 \bigskip
 Over ${\r}$ this question is completely answered. For dimension $4$ and Minkowski signature (see [1] and the references therein) where it is shown  that the only particles with full conformal symmetry are the massless particles with finite helicity. For arbitrary dimension but Minkowski signature it was completely solved by E. Angelopoulos and M. Laoues [2]. We wish to examine this question when ${\r}$ is replaced by a non-archimedean local field $k$ of characteristic $\not=2$.
\bigskip\noindent
{\bf Theorem 2.\/} {\it Massive particles in $V$ do not have conformal symmetry.\/}
\bigskip
If $r\in V$ is a null vector and we consider a massless PUIR $\pi$ of $P=P_k$, the stabilizer of $r$ is the Poincar\' group associated to $V_1$ where $V_1$ is Witt equivalent to $V$ and $\dim (V)-\dim (V_1)=2$. The PUIR $\pi$ is then associated to a PUIR $\pi_1$ of the $k$-Poincar\'e group $P_1$ of $V_1$. It can be shown that if $\pi$ has partial conformal symmetry, then $\pi_1$ has the same property. If $\pi_1$ is massive we stop this process of dimensional reduction and conclude that $\pi$, though massless, has no partial conformal symmetry. Otherwise we continue. This process can be continued till it comes to a stop either at a massive particle or when the corresponding quadratic vector space is anisotropic. In the former case we say the particle is {\it eventually massive.\/}
\bigskip\noindent
{\bf Theorem 3.\/} {\it Eventually massive particles do not have conformal symmetry.\/}
\bigskip\noindent
{\bf Remark.\/} If all the particles defined by the above inductive process are massless we do not know if the original particle has conformal symmetry.
\vskip 0.5 true in
\centerline {\bf References}
\vskip 0.5 true in\item
{[1]} E. Angelopoulos, M. Flato, C. Fronsdal, and D. Sternheimer, {\it Massless particles, conformal group, and De Sitter universe\/}, {\it Phys. Rev.\/}, D23 (1981), 1278--1289.
\medskip\item
{[2]} E. Angelopoulos, and M. Laoues, {\it Masslessness in $n$-dimensions\/}, Rev. Math. Phys., 10(1998), 271--300.
\medskip\item
{[3]} E. G. Beltrametti, {\it Can a finite geometry describe the physical space-time\/}? Universita degli studi di Perugia, Atti del convegno di geometria combinatoria e sue applicazioni, Perugia 1971, 57--62.
\medskip \item
{[4]} E. G. Beltrametti, {\it Note on the $p$-adic generalization of Lorentz transformations\/}, Discrete mathematics, 1(1971), 239--246.
\medskip\item
{[5]} E. G. Beltrametti and Cassinelli, G.,{\it Quantum mechanics and $p$-adic numbers\/}, 2 (1972), 1--7.
\medskip \item
{[6]} C. Carmeli, G. Cassinelli, A. Toigo, and V. S. Varadarajan, {\it Unitary representations of super Lie groups and applications to the classification and multiplet structure of super particles\/}, Comm. Math. Phys., 263 (2006) 217--258.
\medskip\item
{[7]} B. Dragovich, A. Yu. Khrennikov, S. V. Kozyrev, and I. V. Volovich, {\it On $p$-adic mathematical physics\/}, $p$-Adic Numbers, Ultrametric Analysis, and Applications, 1(2009), 1--17.
\medskip \item
{[8]} E. G. Effros, {\it Transformation groups and $C^\ast$-algebras\/}, Ann. Math., 81(1965), 38--55.
 \medskip\item {[9]}
A. N. Kochubei, and M. R. Sait-Ametov, {\it Interaction measures on the space of distributions over the field of $p$-adic numbers\/}, Inf. Dimen. Anal. Quantum Prob. Related Topics, 6(2003), 389-411.
\medskip \item
{[10]} S. Lang, {\it Algebra}, Third Edition, Addison-Wesley, 1993.
\medskip\item
{[11]} G. W. Mackey, {\it Unitary representations of group extensions.\/} I, Acta Math., 99 (1958), 265--311.
\medskip\item
{[12]} G. W. Mackey, {\it Unitary group representations in physics, probability, and number theory\/}, Mathematics Lecture Note
Series, 55. Benjamin/Cummings Publishing Co., Inc., Reading, Mass.,
1978. Second edition. Advanced Book Classics. Addison-Wesley
Publishing Company, Advanced Book Program, Redwood City, CA, 1989.
\medskip \item
{[13]} Yu. I. Manin, {\it Reflections on arithmetical physics\/}, in {\it Conformal invariance and string theory,\/} Acad.
Press, 1989, 293.
\medskip\item
{[14]} C. C. Moore, {\it Group extensions of $p$-adic and adelic linear groups\/}, Publ. Math. IHES, 35 (1968), 5--70.
\medskip\item
{[15]} Y. Nambu, {\it Broken Symmetry} : {\it Selected papers of Y, Nambu\/}, T. Eguchi, and K. Nishijima, eds. World Scientific, River Edge, N.J., 1995.
\medskip\item
{[16]} G. Prasad and M. S. Raghunathan, {\it Topological central extensions of semisimple groups over local fields\/}, Ann. Math., 119 (1984), 143--201; 203--268.
\medskip\item
{[17]} V. S. Varadarajan, {\it Unitary representations of super Lie groups\/}, Lectures given in Oporto, Portugal, July 20--23, 2006.
\medskip\item
{[18]} V. S. Varadarajan, {\it Geometry of Quantum Theory\/}, Second Edition, Springer, 2007.
\medskip\item
{[19]} V. S. Varadarajan, {\it Multipliers for the symmetry groups of $p$-adic space-time\/}, $p$-Adic Numbers, Ultrametric Analysis, and Applications, 1(2009), 69--78.
\medskip\item
{[20]} I. V. Volovich, {\it Number theory as the ultimate theory\/}, CERN preprint CERN-TH.4791/87, 1987; {\it $p$-adic string\/}, Class. Quantum Grav. 4(1987) L83--L87.
\medskip\item
{[21]} E. Wigner, {\it On unitary representations of the inhomogeneous Lorentz group\/}, Ann. Math., 40(1939), 149--204.
\vskip 0.5 true in\noindent
{\mysmall V. S. Varadarajan, Department of Mathematics, UCLA, Los Angeles, CA 90095-1555, USA, {\eightit vsv@math.ucla.edu}} \smallskip\noindent
{\mysmall Jukka Virtanen, Department of Mathematics, UCLA, Los Angeles, CA 90095-1555, USA, {\eightit virtanen@math.ucla.edu}
\medskip

\bye